# Multi-channel SNSPD system with high detection efficiency at telecommunication wavelength

**Shigehito Miki,[1,*] Taro Yamashita,[1] Mikio Fujiwara,[2] Masahide Sasaki,[2] and Zhen Wang[1]**

[1] *Kansai Advanced Research Center, National Institute of Information and Communications Technology, 588-2, Iwaoka, Iwaoka-cho, Nishi-ku, Kobe, Hyogo 651-2492, Japan*

[2] *National Institute of Information and Communications Technology, 4-2-1 Nukui-Kitamachi, Koganei, Tokyo 184-8795, Japan*

[*]*Corresponding author: s-miki@nict.go.jp*

We developed a four-channel superconducting nanowire single-photon detector system based on a Gifford-McMahon cryocooler. All channels showed a system detection efficiency (at a 100 Hz dark-count rate) higher than 16% at 1550 nm wavelength, and the best channel showed a system DE of 21% and 30% at 1550 nm and 1310 nm wavelength, respectively.

In recent times, multichannel superconducting nanowire single-photon detector (SNSPD [1]) systems based on closed-cycle cryocoolers have been recognized as promising instruments in the field of optical quantum information technology; this is because SNSPDs deliver a good performance and are capable of continuous and stable operation without any liquid cryogen. At present, the type of SNSPDs mostly used in the multichannel systems are single-layer nanowire devices. They typically showed a practical system detection efficiency (DE) of 1-3 % at 1550 nm wavelength, low dark-count rate (DCR) of 10-100 Hz, and excellent timing jitter of 30-100 ps [2,3]. Although they have been successfully employed in quantum key distribution (QKD) [2,4-6], further improvement in the system performance, especially in the system DE, is desired.

An effective method of improving the system DE is enhancement of the photoabsorption coefficient by integrating an optical cavity structure with the SNSPD device (OC-SNSPD)[7]. Moreover, efficient optical coupling to the meander nanowire area simultaneously is crucial, and a primary concern is how to implement the OC-SNSPDs in practical multichannel system. Although the successful implementation of a single OC-SNSPD in a practical closed-cycle cryocooler systems has been reported [8,9], the development of multichannel systems has not yet been reported. Since the BB84 [10] protocol, which is currently a most matured protocol with complete security certification, requires four channels at the receiver side [5], it is necessary to have four channels in the system for use in these applications. In this letter, we report the development of a practical four-channel OC-SNSPD system with higher system DE than 16 % at a wavelength of 1550 nm. We present a newly developed compact fiber-coupled packaging technique using lenses and describe their system DE and DCR performances.

The devices used in this study were patterned onto 4-nm-thick NbN thin films on MgO substrates [11]. We fabricated 80-nm-wide NbN meander nanowires covering an area of 15 × 15

µm$^2$ with a filling factor of 62.5%. The superconducting critical temperature T$_c$ and critical current density J$_c$ of nanowires were 10.2-10.5 K and 4-7 × 10$^{10}$ A/m$^2$, respectively. An optical cavity structure consisting of a 100-nm-thick Au mirror and a 250-nm-thick SiO cavity were covered on the meander nanowire area. The thicknesses of Au mirror and SiO cavity were designed for wavelengths of 1300-1600 nm.

Figure 1 (a) shows the schematic layout of the fiber-coupled packaging for OC-SNSPDs. This compact fiber-coupled packaging technique was modified from the one used for single layer SNSPD [2,3], which is simple and has high reliability. A fiber ferrule was fixed to the fiber-holding block in advance by using an adhesive so that the distance from the exit end to the rear surface of the OC-SNSPD chip was 20 µm at low temperature. OC-SNSPD chips were mounted on chip-mounting blocks which had a through hole at the center of the chip-mounting area. An MU-type fiber ferrule was inserted through this hole from the rear. The fiber-holding block was joined to the chip-mounting block from the rear, and the two blocks were accurately aligned so that the incident light spot illuminated the center of the meander area. The dimensions of the packaging blocks which could be used for OC-SNSPDs are almost of the same size (15 mm (length) × 15 mm (width) × 10 mm (thickness)) as those used for single-layer SNSPD chips, as shown in fig.1 (b). Hence, OC-SNSPD packaging blocks were installed in the multichannel Gifford-McMahon (GM) cryocooler system without any modification, which can simultaneously cool six SNSPD packages to 2.9 K with a thermal fluctuation range of ±10 mK [3].

To achieve efficient optical coupling, light beam waist on the meander nanowire area must be smaller than the size of the nanowire area. Since the OC-SNSPDs have to be illuminated from the rear side through the substrate, small-gradient index (GRIN) lenses were used to reduce the beam waist at a distant from exit-end. In order to embed lenses into the compact packages,

GRIN lenses with a diameter of 125 μm, which is equal to the clad diameter of a single-mode (SM) optical fiber, are directly fusion spliced to the end of the optical fiber. Since the fiber-spliced lenses were inserted into the MU fiber ferrule, the shape of the end of fiber did not change at all from that without lenses, as shown in Fig.1 (c). The numerical aperture and length of the two lenses are chosen so that the focal length is equal to the appropriate distance in the packaging and the beam waist becomes as small as possible. As a result, the beam waist ($2\omega_0$) was estimated to be 8–10 μm on the meander nanowire area, when the distance between the exit-end and the substrate is 20 μm and the thickness of the MgO substrate is 400 μm. This beam waist is sufficiently small to allow efficient optical coupling with the meander nanowire area of $15 \times 15$ μm$^2$.

To verify the effectiveness of the fiber-spliced GRIN lenses, we measured the system DE versus the DCR of the OC-SNSPD device, with and without the lenses, as shown in Figure 2. Here, system DE is defined as the ratio of the output count rate and the input photon flux to the system. The system DE (at a DCR of 100 Hz) of a device without lenses remained to be a small value of 2.8%; this was because the beam waist expanded on the meander nanowire area, thereby resulting in poor optical coupling. With focusing by GRIN lenses, system DE was much improved to be 21%. Figure 3(a) shows system DE versus DCR and Fig. 3(b) shows the system DE and DCR versus bias current normalized by superconducting critical current $I_c$ at wavelengths of 1310 nm and 1550 nm. A maximum system DE reached 40% and 28%, at 1310 nm and 1550 nm wavelength, respectively, at the DCR of several thousand Hz, where the bias current was just below $I_c$ (~0.99$I_c$). This implies the optical cavity also worked efficiently at 1310 nm wavelength and the device DE at 1310 nm became higher than that at 1550 nm.

By using compactly packaged OC-SNSPDs, we built up a four-channel SNSPD system. Figure 4 shows the system DE versus DCR of the four channel OC-SNSPDs at a wavelength of 1550 nm. Needless to say, all the channels can operate simultaneously without any time gating, and hence, our OC-SNSPD can be used in QKD protocols and other quantum optical applications. The system DE of all the channels show similar dependencies on the DCR and exceed 16% and 20% at DCRs of 100 Hz and 2000 Hz respectively, which are significantly higher than those of standard devices [4]. Although the system DE varies from channel to channel, we believe values of all the channels can exceed those achieved in the present study by improving the yield of nanowire uniformity. The performance of the packaged OC-SNSPDs do not show any significant change during several thermal cycles. These results indicate that our packaging technique exhibits a high stability against thermal cycling.

As a future study, an effective optical coupling to smaller devices should be considered because miniaturization of device size are certainly effective for further improvement of both device DE and response speed [12,13]. It should be noted that this will be possible by utilizing a combination of fiber-spliced GRIN lenses and substrate thickness-reduction technique [9]. For example, if the thickness of the MgO substrate is reduced to 50 μm and the GRIN lenses are optimally designed, the beam waist can be reduced to 4-5 μm, thus making it possible to achieve efficient optical coupling to smaller devices.

In conclusion, we have developed a four-channel SNSPD system using OC-SNSPD devices. To achieve efficient optical coupling in the multichannel system, a compact fiber-coupled packaging technique that employ GRIN lenses has been developed. The beam waist on the meander nanowire area has been successfully reduced for achieving effective optical coupling to devices with an area of $15 \times 15$ μm$^2$. All the channels showed a system DE higher than 16%, and best device showed 21% at a DCR of 100 Hz and wavelength of 1550 nm. These DE values are significantly higher than those of standard multi-channel SNSPD systems, and definitely make a great impact to the QKD and various applications.

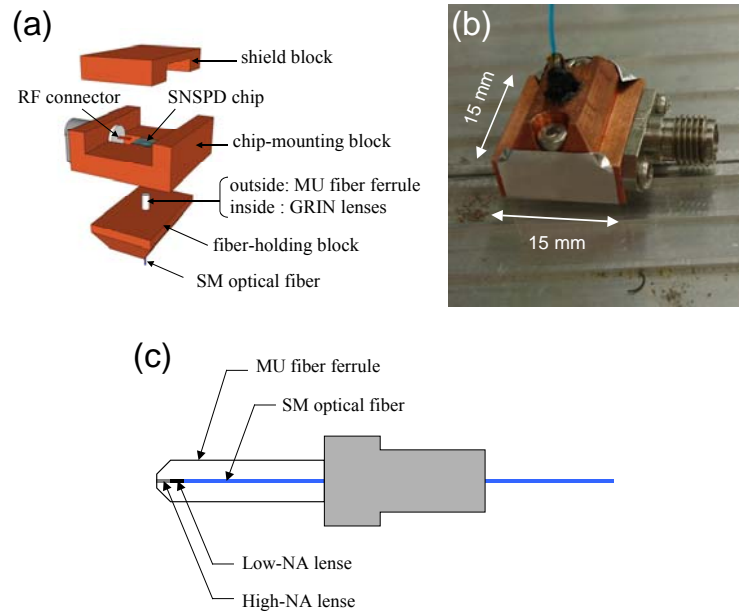

Fig. 1. (a) Schematic layout and (b) photograph of fiber-coupled packaging for OC-SNSPDs (c) configuration of GRIN lenses connected to optical fiber.

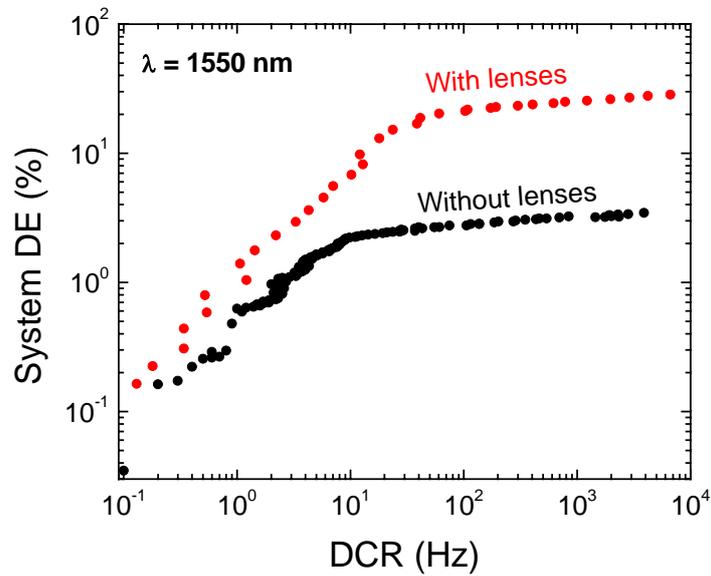

Fig.2. System DE versus DCR of OC-SNSPD device packaged with and without GRIN lenses. The System DE (at a 100 Hz DCR) reached a level of 21% by using GRIN lenses

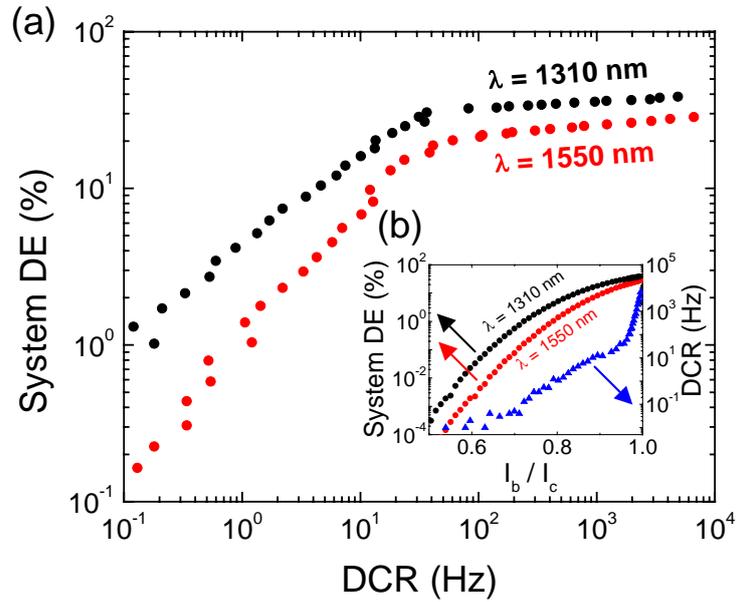

Fig.3. The best performance of our OC-SNSPD device. (a)System DE versus DCR and (b) system DE and DCR versus bias current of OC-SNSPD device at 1310 nm and 1550 nm wavelength.

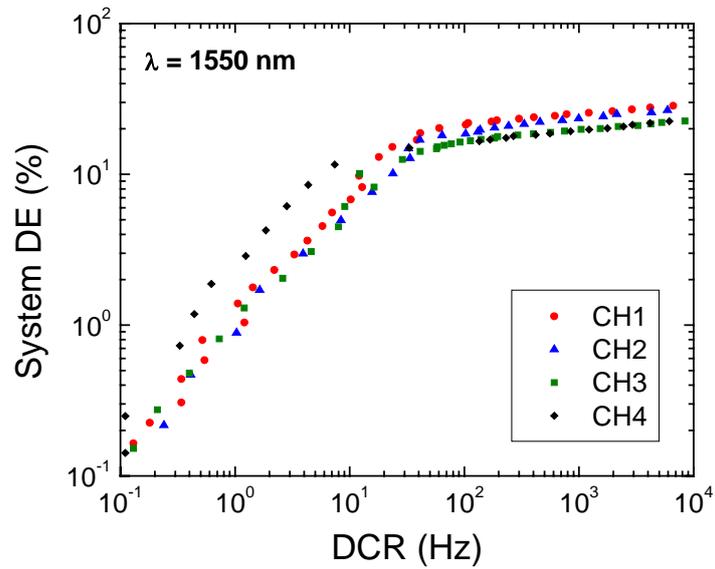

Fig.4. System DE versus DCR of four-channel OC-SNSPDs at a wavelength of 1550 nm.